\documentclass[a4paper]{jpconf}
\usepackage{graphicx}
\newcommand{\be}{\begin{equation}}
\newcommand{\ee}{\end{equation}}
\newcommand{\bea}{\begin{eqnarray}}
\newcommand{\eea}{\end{eqnarray}}
\newcommand{\nn}{\nonumber}
\usepackage{amstext,amssymb}
\usepackage{amsmath}

\def\U1mt{U(1)_{L_\mu-L_\tau}}

\def\ol{\overline}
\def\nl{\nonumber\\}
\usepackage{slashed}
\global\long\def\d{\partial}

\begin{document}
\title{Exploring Dark Matter, Neutrino mass and flavour anomalies  in  $L_{\mu}-L_{\tau}$ model}
\author{Rukmani Mohanta$^1$, Shivaramakrishna Singirala$^2$, Suchismita Sahoo$^3$}
\address{$^1$School of Physics, University of Hyderabad, Hyderabad - 500046, India}
\address{$^2$Discipline of Physics, Indian Institute of Technology Indore, Indore-453552, India}
\address{$^3$Theoretical Physics Division, Physical Research Laboratory, Ahmedabad-380009, India}

\ead{$^1$rukmani98@gmail.com}

\begin{abstract}
We investigate Majorana dark matter in a new variant of 
 $U(1)_{L_{\mu}-L_{\tau}}$ gauge extension of Standard Model, containing three additional neutral fermions $N_{e}, N_{\mu}, N_{\tau}$, along with a $(\bar{3},1,1/3)$ scalar Leptoquark (SLQ) and an inert doublet, to study the phenomenology of dark matter, neutrino mass generation and flavour anomalies on a single platform. 
The lightest mass eigenstate of the $N_{\mu}, N_{\tau}$ neutral fermions plays the role of dark matter. We compute the WIMP-nucleon cross section in leptoquark portal and the relic density mediated by inert doublet components, leptoquark and the new $Z^{\prime}$ boson. We constrain the parameter space consistent with Planck limit on relic density, PICO-60 and LUX bounds on spin-dependent direct detection cross section. We also  discuss about the neutrino mass generation at one-loop level and the viable parameter region to explain current neutrino oscillation data.
 The $Z^\prime$ gauge boson of extended $U(1)$ symmetry and the SLQ   play an important role in settling the known issues of flavor sector.   
\end{abstract}

\section{Introduction}
Though the Standard Model (SM) is quite successful in explaining all the observed  data so far, it fails to provide satisfactory answer to some of the open issues like  matter-antimatter asymmetry,  neutrino mass, nature dark matter (DM) and dark energy, thus necessitates the existence of new physics (NP) beyond it. So far no direct evidence of NP has been established  either in energy frontier or intensity frontier. However, in recent times several discrepancies at the level of $(2-4)\sigma$ have been observed in the rare $B$ decay modes,  mediated by $b \to s \ell \ell$ transitions. These include  the famous  $P_5^\prime$ observable in the decay distribution of  $\bar B \to \bar K^* \mu \mu$ mode \cite{BtoKstar},   the lepton nonuniversality (LNU) ratio, $R_K \equiv {\rm Br}(B \to K \mu \mu)/ {\rm Br}(B \to K ee)=0.846^{+0.060~+0.0016}_{-0.054~-0.014}$ \cite{LNU-RK}, which shows a deviation of $2.5\sigma$ from its SM result, and the recent measurements on $R_{K^*}$ ratio by LHCb Collaboration \cite{LNU-RKstar}
has also $2.2\sigma~(2.4\sigma)$ discrepancy from the respective SM predictions \cite{RKstar-SM} in $q^2\in [0.045,1.1]~([1.1,6])$ bins, respectively. In this work, we would like to consider the anomaly free  $U(1)_{L_\mu-L_\tau}$ gauge extension of the SM  along with a scalar leptoquark (SLQ) to address the flavour anomalies, Neutrino oscillation and Dark Matter in a single platform.

The layout of the article is as follows. In Sec. 2, we briefly outline the model description. Sec. 3 (4) contains the discussion on  neutrino mass generation and the parameter space consistent with  DM (flavor) study and  our results are summarized in Sec. 5. 

\section{New variant of $L_{\mu}-L_{\tau}$ model with a scalar leptoquark}

We consider the  $U(1)_{L_{\mu}-L_{\tau}}$ \cite{He:1990pn} extension of SM, with three additional neutral fermions $N_{e}, N_{\mu}, N_{\tau}$ having $L_{\mu}-L_{\tau}$ charges $0,1$ and $-1$ respectively. The scalar content of our model includes  a scalar singlet $\phi_2$, an inert doublet $\eta$ and a scalar leptoquark  $S_{1}(\bar{3},1,1/3)$.  We impose an additional $Z_2$ symmetry under which all the new fermions, $\eta$ and the leptoquark are odd and rest are even. The additional particle content of our model and their corresponding charges are displayed in Table. \ref{mutau_model}\,.
\begin{table}[htb]
\begin{center}
\begin{tabular}{|c|c|c|c|c|}
	\hline
			& Field	& $ SU(3)_C \times SU(2)_L\times U(1)_Y$	& $U(1)_{L_{\mu}-L_{\tau}}$	& $Z_2$\\
	\hline
Fermions			& $N_{e}$						& $(\textbf{1},\textbf{1},~   0)$	&  $0$	& $-$\\
			& $N_{\mu}$						& $(\textbf{1},\textbf{1},~   0)$	&  $1$	& $-$\\
			& $N_{\tau}$						& $(\textbf{1},\textbf{1},~   0)$	&  $-1$	& $-$\\
	\hline
	Scalars	& $H$							& $(\textbf{1},\textbf{2},~ 1/2)$	&   $0$	& $+$\\
		& $\eta$							& $(\textbf{1},\textbf{2},~ 1/2)$	&   $0$	& $-$\\
			& $\phi_2$						& $(\textbf{1},\textbf{1},~   0)$	&  $2$	& $+$\\  
			& $S_1$						& $(\bar{\textbf{3}},\textbf{1},~   1/3)$	&  $-1$	& $-$\\    
	\hline
	\hline
\end{tabular}
\caption{New fields and their charges of the proposed $U(1)_{L_{\mu}-L_{\tau}}$ model.}
\label{mutau_model}
\end{center}
\end{table}

The Lagrangian of the present model can be written as
\begin{align}
{\cal L} &={\cal L}_{\rm SM}  -{1 \over 4} Z'_{\mu\nu} Z^{'\mu\nu} -g_{\mu\tau}\overline{\mu}_L \gamma^\mu \mu_L Z_\mu^\prime 
            - g_{\mu\tau} \overline{\mu}_R \, \gamma^\mu \mu_R Z_\mu^\prime  +g_{\mu\tau}\overline{\tau}_L  \gamma^\mu \tau_L Z_\mu^\prime 
           + g_{\mu\tau} \overline{\tau}_R \, \gamma^\mu \tau_R  Z_\mu^\prime \nl
            &+\overline{N}_{e} i \slashed{\d}\,N_{e}+ \overline{N}_{\mu} \left(i \slashed{\d} - g_{\mu\tau} \,Z_\mu^\prime \gamma^\mu \right)N_{\mu}	 +  \overline{N}_{\tau} \left(i \slashed{\d} + g_{\mu\tau} \,Z_\mu^\prime \gamma^\mu \right)N_{\tau}  - \frac{f_\mu}{2}\left({\ol{N_{\mu}^c}} N_{\mu}\phi_2^{\dagger}+{\rm h.c.}\right) \nl &- \frac{f_\tau}{2}\left({\ol{N_{\tau}^c}} N_{\tau}\phi_2 + {\rm h.c.} \right) -\frac{1}{2}M_{ee}\ol{N_{e}^c} N_{e} -\frac{1}{2}M_{\mu\tau}(\ol{N_{\mu}^c} N_{\tau} + \ol{N_{\tau}^c} N_{\mu}) - \sum_{q=d,s,b} (y_{q R}\; \ol{d_{qR}^c} S_1 N_{\mu} + {\rm{h.c.}})  \nl
&- \sum_{l=e,\mu,\tau} (Y_{\beta l} (\ol{\ell_L})_\beta \tilde \eta N_{lR} + {\rm{h.c}})+ \left|\left(i \d_\mu - \frac{g}{2} \boldsymbol{\tau}^a\cdot\bold{W}_\mu^a  -\frac{g^{\prime}}{2}B_\mu\right) \eta \right|^2 
+ \left| \left(i \d_\mu -\frac{g^{\prime}}{3}B_\mu + g_{\mu\tau} \,Z_\mu^\prime  \right) S_1\right|^2
\nl &
+ \left| \left(i \d_\mu -2 g_{\mu\tau} \,Z_\mu^\prime  \right) \phi_2\right|^2 
- V(H,\eta,\phi_2,S_1).
\label{eq:Lag}
\end{align}
where the scalar potential  $V(H,\eta,\phi_2,S_1)$ is expressed  as
\begin{align}
&V(H,\eta,\phi_2,S_1) =  \mu^2_H  H^\dagger H + \lambda_H (H^\dagger H)^2  + \mu_{\eta}(\eta^{\dagger}\eta) + \lambda_{\eta}(\eta^{\dagger}\eta)^2 + \frac{\lambda''_{H\eta}}{2}\left[(H^{\dagger}\eta)^2 + {\rm h.c.}\right]\nn\\
&~~~~~~~+ \lambda'_{H\eta}(H^{\dagger}\eta)(\eta^{\dagger}H)     
      + \mu^2_{2} (\phi^\dagger_2 \phi_2) + \lambda_{2} (\phi^\dagger_2 \phi_2)^2 
      +\mu^2_{S} ({S_1}^\dagger {S_1})  +\lambda_{S} ({S_1}^\dagger {S_1})^2 +  \lambda_{S2}(\phi^\dagger_2 \phi_2) ({S_1}^\dagger {S_1})\nn\\
&~~~~~~~       +\lambda_{\eta2}(\phi^\dagger_2 \phi_2) (\eta^\dagger \eta) 
      +  \left[\lambda_{H2} (\phi^\dagger_2 \phi_2) + \lambda_{HS} (S^\dagger_1 S_1+\lambda_{H\eta}(\eta^{\dagger}\eta))\right](H^\dagger H) + \lambda_{S\eta}({S_1}^\dagger S_1) (\eta^\dagger \eta).\nn
\label{eq:potential}
\end{align}
The $U(1)_{L_\mu -L_\tau}$ symmetry is spontaneously broken when the scalar field $\phi_2$ acquires the  VEV $v_2$. Then the SM Higgs doublet breaks the SM gauge group to low energy theory. The fermion and scalar mass matrices take the form
\begin{align}
	M_N
	=
	\begin{pmatrix}
		 \frac{1}{\sqrt{2}}f_{\mu}v_2	& M_{\mu\tau}	\\
		 M_{\mu\tau}	& \frac{1}{\sqrt{2}}f_{\tau}v_2				\\
	\end{pmatrix} ,
    \quad M_S
	=
	\begin{pmatrix}
		 2 \lambda_H v^2   & {\lambda}_{H2} {v}v_2  \\
 {\lambda}_{H2} {v}v_2  & 2 \lambda_{2} v^2_2				\\
	\end{pmatrix} .
\end{align}
One can diagonalize these mass matrices by using the $2\times 2$ rotation matries as $U_{\alpha(\zeta)}^T M_{N(S)} U_{\alpha(\zeta)} = {\rm{diag}}~[M_{N_{-}(H_{1})},M_{N_{+}(H_{2})}]$. The lightest mass eigenstate $N_-$ acts as the DM candidate and the scalar field $H_1$ is considered to be the observed Higgs with  mass $M_{H_1} = 125$ GeV. The neutrino masses generated at one-loop level,  given as
\begin{equation}
({\cal M}_\nu)_{\beta\gamma} = {\lambda''_{H\eta} v^2  \over16 \pi^{2}} 
\sum_{l=e,\mu,\tau} {Y_{ \beta l} Y_{\gamma l} M_{Dl} \over m_0^{2} - M^{ 2}_{Dl}} \left[ 
1 +{M^{ 2}_{Dl} \over m_0^{2}-M^{ 2}_{Dl}} \ln { M^{ 2}_{Dl}\over m_0^{2}  }  \right].\label{nu-mass}
\end{equation}
Here $M_{Dl} = (U^T M_N U)_{l} = {\rm diag}(M_{ee},M_-,M_+)$.

\section{Dark matter phenomenology}
\begin{figure}[htb]
\centering
\includegraphics[scale=0.29]{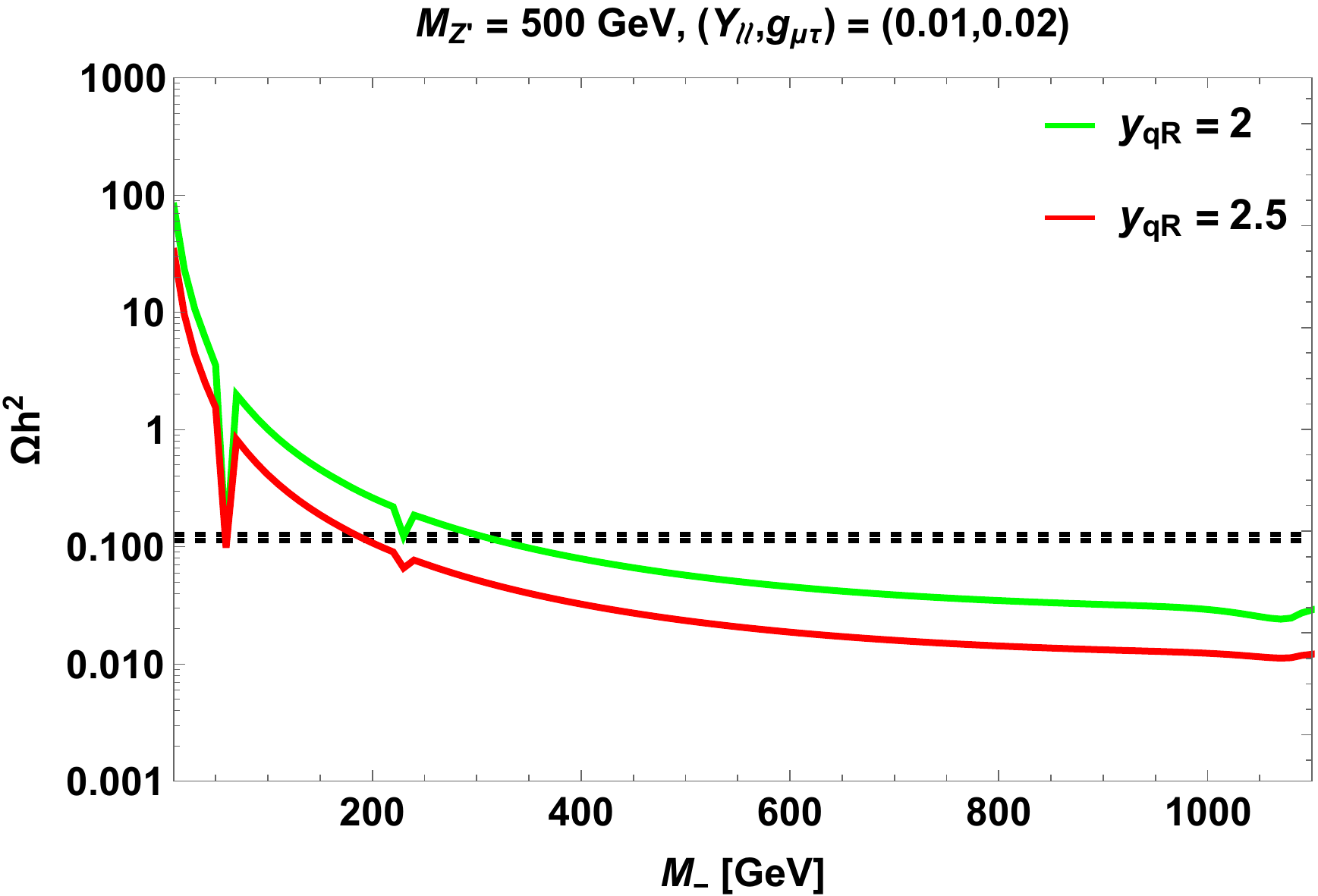}
\quad
\includegraphics[scale=0.29]{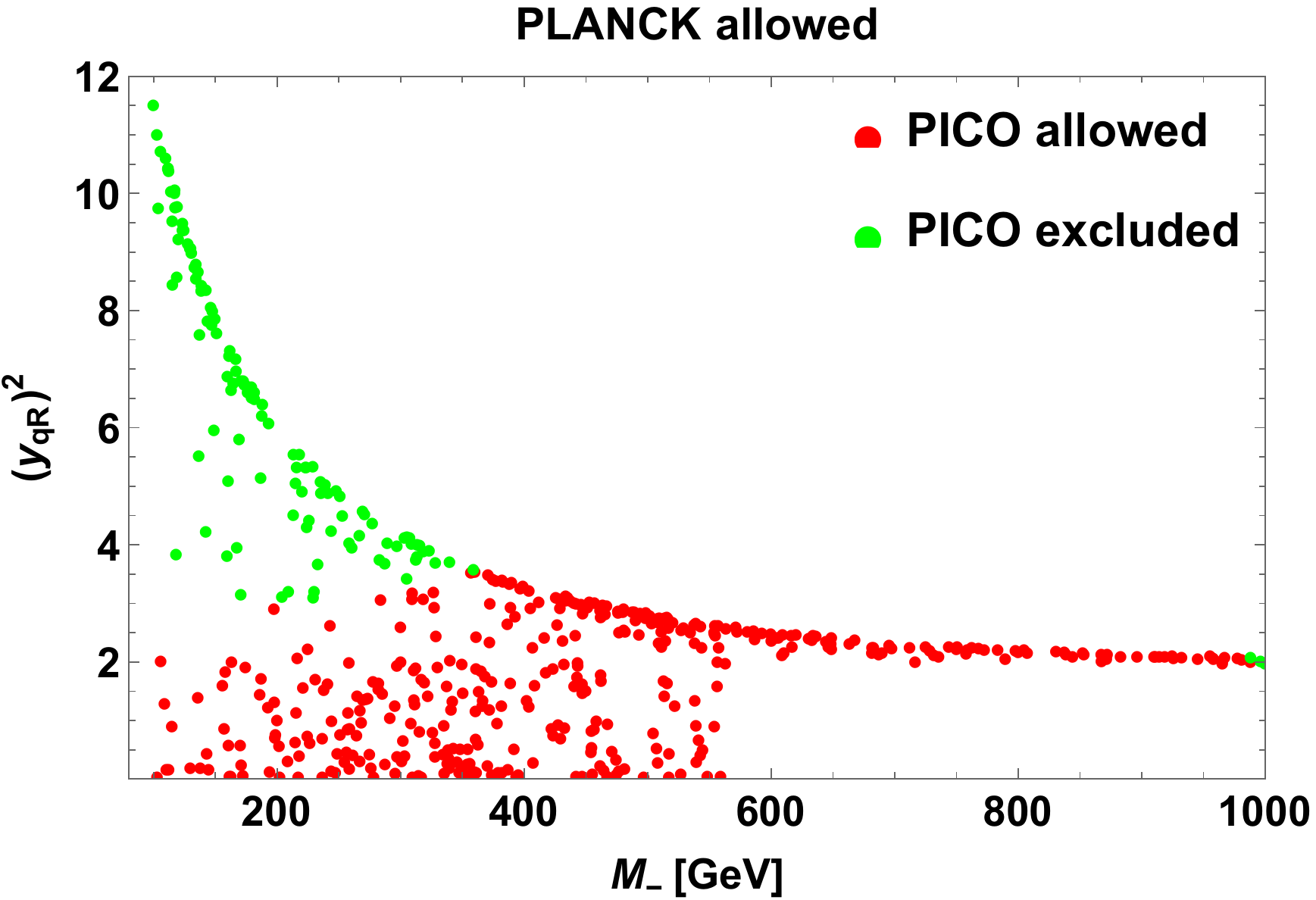}
\caption{Left panel depicts the relic density as a fucntion of DM mass for benchmark values of model parameters $(M_{S_1},M_{\eta^+}, M_{\eta_{e,o}},M_{H_2}) = (1.2,2,1.5,2.2)$ TeV. Horizontal dashed lines correspond to Planck limit \cite{Aghanim:2018eyx}. Right panel shows the paramter space consistent with $3\sigma$ range of Planck central value.}\label{relic_curve}
\end{figure}

{\bf A. Relic density}\\
The annihilations include s-channel processes mediated by $H_1,H_2$ with $f\bar{f}, W^+W^-,ZZ,Z^\prime Z^\prime, H_iH_j$ in the final state, $Z^\prime$ portal giving $Z^\prime H_i$, pair of leptons of $\mu$ and $\tau$ type in output. Finally, in $S_1$-portal $d\bar{d},s\bar{s},b\bar{b}$ arise as output and in $\eta$-portal, pair of charged and neutral leptons arise in the final state. The model parameters to be analyzed are $(M_-,g_{\mu\tau},M_{Z^\prime},Y_{\beta l},y_{qR})$. Left panel of Fig. \ref{relic_curve} shows relic density with DM mass for various values of Yukawa couplings.

{\bf B. Direct searches}\\
In the scalar portal, one can obtain contribution from spin-dependent (SD) interaction mediated by SLQ, of the form $\overline{N_-}\gamma^\mu\gamma^5 N_- \overline{q}\gamma_\mu\gamma^5$. The corresponding cross section is given by \cite{Agrawal:2010fh}
\begin{equation}
\sigma_{\rm SD} = \frac{ \mu_r^2}{\pi} \frac{\cos^4\alpha}{(M_{S_1}^2 - M_-^2)^2}\left[y_{dR}^2\Delta_d + y_{sR}^2\Delta_s\right]^2 J_n(J_n+1),
\end{equation}
where, $J_n = \frac{1}{2}$, $\mu_r = \frac{M_- M_n}{M_-+M_n}$ with $M_n \simeq 1$ GeV for nucleon.  Right panel of Fig. \ref{relic_curve}, depicts the parameter space consistent with $3\sigma$ range of Planck limit, with green data points violate the PICO-60 bound \cite{Amole:2017dex}.
\section{Flavor phenomenology}

\begin{figure}
\centering
\includegraphics[scale=0.29]{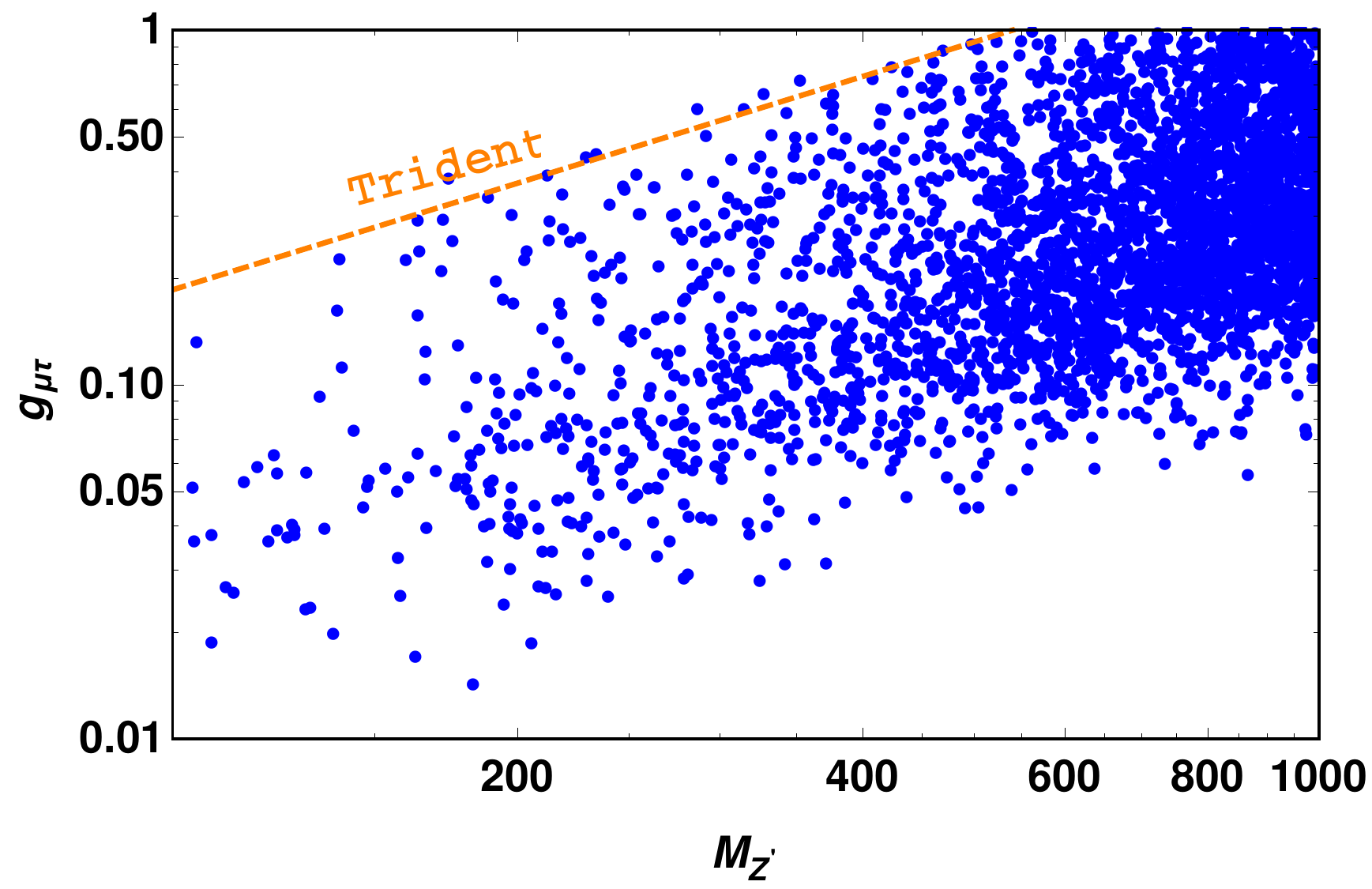}
\quad
\includegraphics[scale=0.24]{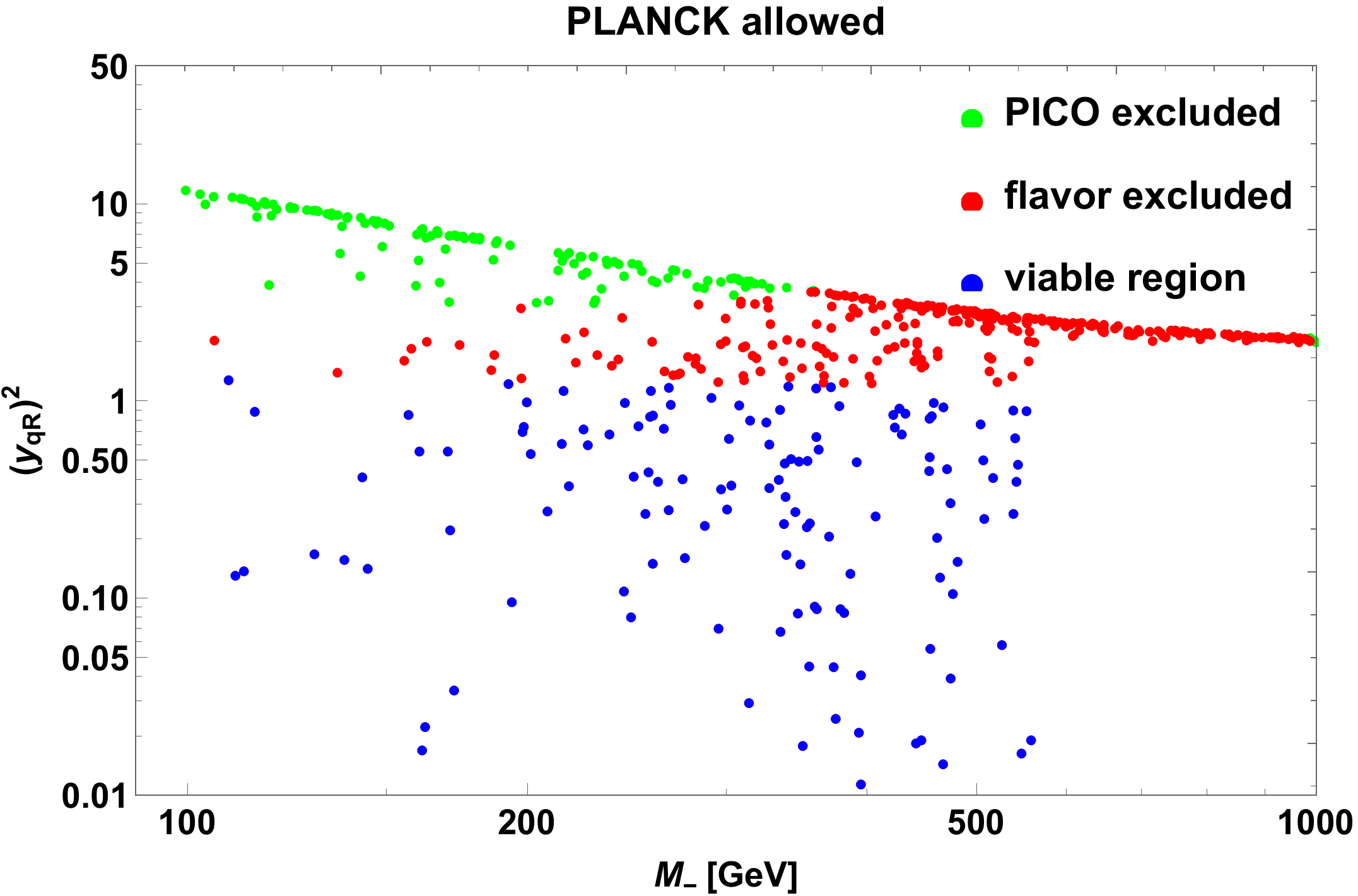}
\caption{Constraints on $M_{Z^\prime}-g_{\mu\tau}$ (left panel) and $M_--y_{qR}^2$ (right panel) planes.}\label{Fig:Con-Flavor}
\end{figure}

After obtaining  the allowed parameter space  from DM study, we further constrain the new parameters from the present experimental limits on ${\rm Br}(\tau \to \mu \nu_\tau \bar \nu_\mu)$, ${\rm Br}( B \to X_s \gamma)$,  ${\rm Br}(\bar B^0 \to \bar K^0 \mu^+ \mu^-)$, ${\rm Br}(B^+ \to K^+ \tau^+ \tau^-)$ and $B_s-\bar{B_s}$ mixing.  Comparing the theoretical predictions with their respective $3\sigma$ experimental data, we constrain the allowed parameter space of $M_{Z^\prime}-g_{\mu\tau}$ (left) and $M_--y_{qR}^2$ (right), presented in Fig. \ref{Fig:Con-Flavor}\,. 
In the right panel, the blue points represent the allowed space obtained from both DM and flavor observables (DM+Flavor). Both the red and blue points represent the constraints from only DM study, which are denoted as DM-I and DM-II regions.  

\begin{figure}[htb]
\centering
\includegraphics[scale=0.29]{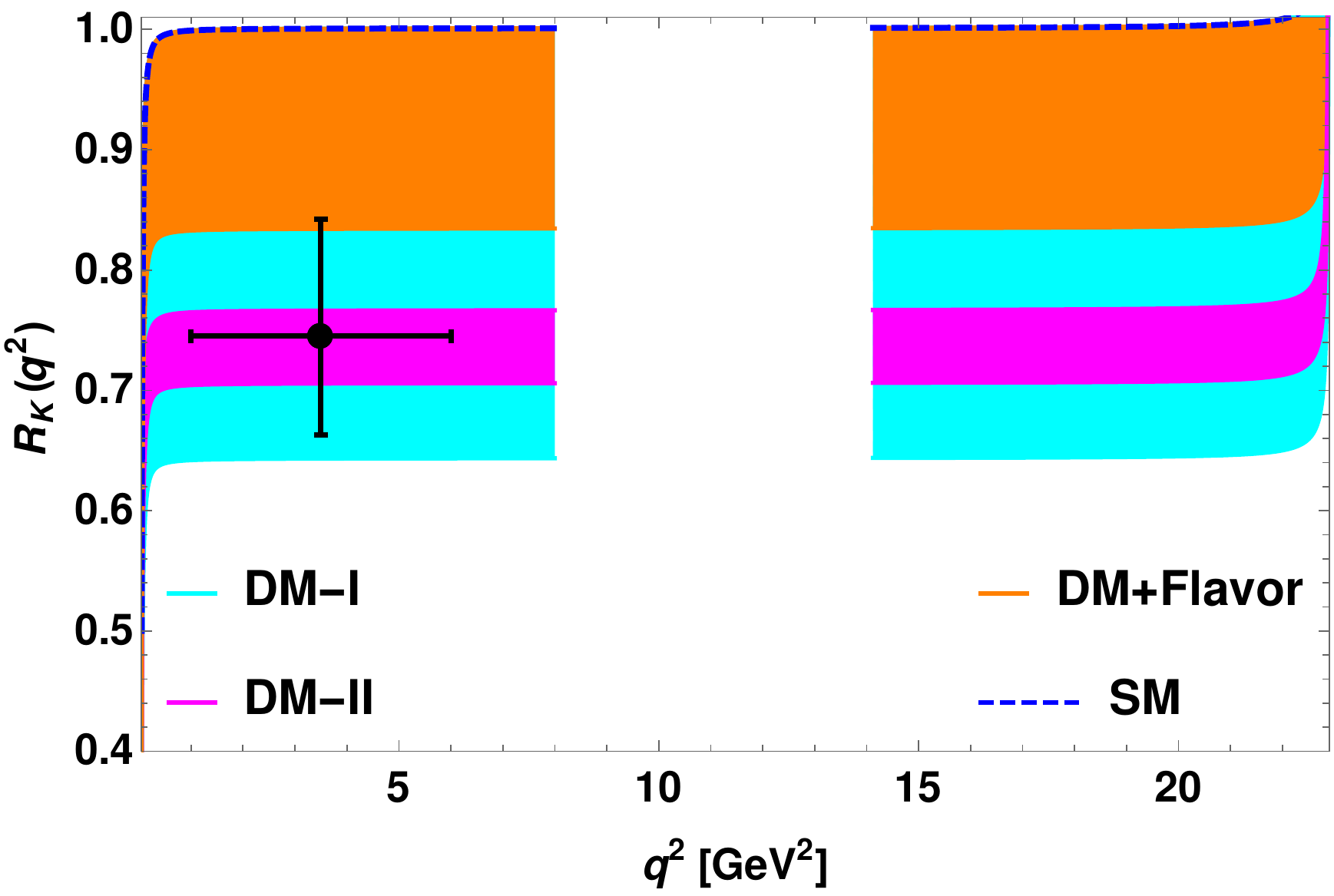}
\quad
\includegraphics[scale=0.4]{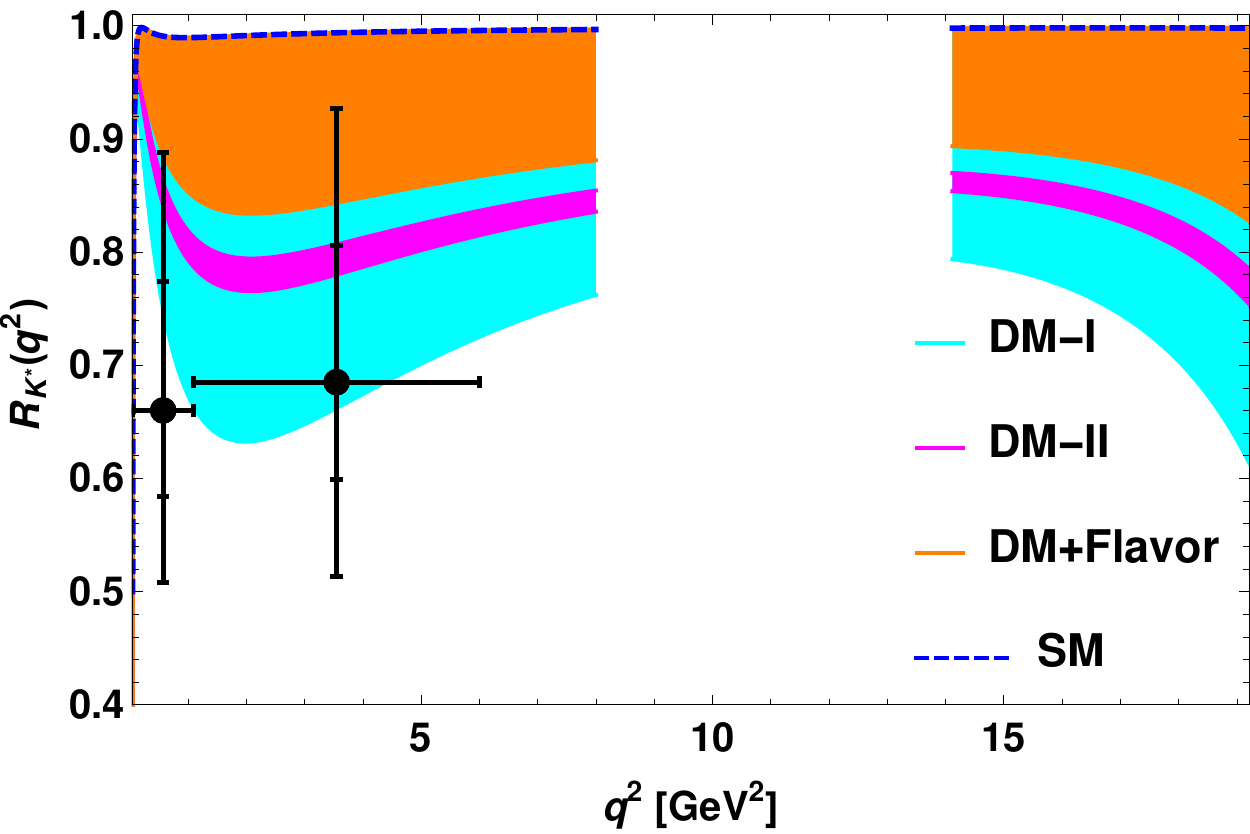}
\caption{The $q^2$ variation of $R_K$ (left), $R_{K^*}$ (right) panel.  }\label{Fig:result}
\end{figure}

We now discuss the effect of constrained new parameters obtained from DM and flavor phenomenology on the lepton nonuniversality ratios  $R_{K^{(*)}}$.  The $q^2$ variation of $R_K$ (top-left) and $R_{K^*}$ (top-right)  ratios are shown in Fig. \ref{Fig:result} in the full $q^2$ region excluding the intermediate  dominant $c\bar{c}$ resonance region. Here blue dashed lines stand for SM predictions, cyan (magenta) bands represent the DM-I (DM-II) contributions and the orange bands are obtained by using the DM+Flavor allowed parameter space. The experimental data are presented in black color \cite{LNU-RK, LNU-RKstar}. We observe that though the LNU ratios deviate significantly from their SM predictions for all the  regions of allowed parameter space, the constraint obtained from DM observables comparatively has larger impact on these ratios. 

\section{Conclusion}
We  scrutinized the  dark matter, neutrino mass and flavor anomalies in the $U(1)_{L_\mu-L_\tau}$ extended SM. We constrained the new parameters from both the dark matter and flavor phenomenology. We then checked the effects of allowed parameter space on $R_K$ and $R_{K^{*}}$. We found that the constraint obtained from only DM study show comparatively good impact than the flavor observables. 

{\bf Acknowledgments}
RM would like to acknowledge the financial support received from SERB, Govt. of India through grant nos. ITS/2019/003369 and EMR/ 2017/001448.

\end{document}